# Droplet morphology-based wettability tuning and design of fog harvesting mesh to minimize mesh-clogging

Arani Mukhopadhyay[1†], Arkadeep Datta[†], Partha Sarathi Dutta[1], Amitava Datta, Ranjan Ganguly*

Advanced Materials Research and Applications (AMRA) Laboratory
Department of Power Engineering, Jadavpur University, Kolkata-700106, India
[†]These authors contributed equally
*Corresponding author email: ranjan.ganguly@jadavpuruniversity.in

**Abstract:**

Fog harvesting relies on intercepting atmospheric or industrial fog by placing a porous obstacle, e.g., a mesh and collecting the deposited water. In the face of global water scarcity, such fog harvesting has emerged as a viable alternative source of potable water. Typical fog harvesting meshes suffer from poor collection efficiency due to aerodynamic bypassing of the oncoming fog stream and poor collection of the deposited water from the mesh. One pestering challenge in this context is the frequent clogging up of mesh pores by the deposited fog water, which not only yields low drainage efficiency but also generates high aerodynamic resistance to the oncoming fog stream, thereby negatively impacting the fog collection efficiency. Minimizing the clogging is possible by rendering the mesh fiber superhydrophobic, but that entails other detrimental effects like premature dripping and flow-induced re-entrainment of water droplets into the fog stream from the mesh fiber. Herein, we improvise on the traditional interweaved metal mesh designs by defining critical parameters, viz., mesh pitch, shade coefficient, and fiber wettability, and deduce their optimal values from numerically and experimentally observed morphology of collected fog-water droplets under various operating scenarios. We extend our investigations over a varying range of mesh-wettability, including superhydrophilic and hydrophobic fibers, and go on to find optimal shade coefficients which would theoretically render *clog-proof* fog harvesting meshes. The aerodynamic, deposition, and overall collection efficiencies are characterized. Hydrophobic meshes with square pores, having fiber diameters smaller than the capillary length scale of water, and an optimal shade coefficient, are found to be the most effective design of such *clog-proof* meshes.

1: Present address: Department of Mechanical and Industrial Engineering, University of Illinois at Chicago, IL, 60607, USA

# 1 Introduction

The accentuating crisis of freshwater in the last few decades has led to a serious drive for the development of technology focused on sustainable water harvesting [1]. Harvesting fog from diverse sources such as natural environments [2] or industrial cooling towers [3] using a mesh has emerged as an affordable and viable alternative. Polypropylene fiber-based Rachel meshes, featuring trapezoidal pores have been extensively used for atmospheric fog harvesting in different parts of the world [4, 5]. However, for heavy fogging under high stream velocities, as is typical in industrial scenarios, sturdy metal meshes having square pores are better suited for long-term use [6]. A fog-laden stream passing through a mesh deposits the fog droplets (ranging from 4 to 30 µm in diameter [7, 8]) on the fibers by inertial impaction, physical interception, and Brownian diffusion [9]. Progressive deposition of these fog-droplets and subsequent coalescence of the deposited liquid lead to the formation of bigger droplets ($O$ (~1 mm)) of different shapes on the fibers. These growing droplets may eventually touch the neighboring mesh fibers, when they clog the mesh pore either partly or completely (see Figure 1A). A clogged mesh pore offers a greater aerodynamic resistance to the oncoming fog stream, which in turn diminishes the fog droplet deposition. The droplet shape on a fiber depends on its volume, the wettability of the mesh fiber and the fiber diameter [10, 11]. Therefore, the morphology of the droplet and its relative size with respect to the mesh dimensions, and the fiber wettability play important roles [12] in ascertaining the fog-harvesting performance of a mesh.

Wettability-engineering has been extensively leveraged by several researchers to enhance fog harvesting on sturdy impervious flat surfaces by advocating the use of 3D features like cascading patterns [13], bumps [14] or bio-inspired surface modifications [15, 16, 17, 18] and micro-/nano-scale surface embellishments [19] for enhancing water capture. However, these impervious plates suffer from significant aerodynamic bypassing – the fog stream gets diverted by the plate in the fog-flow path [20] – leading to minimal fog deposition. Biomimetic surfaces, replicating the Namib desert beetle *onymacris unguicularis* [21, 22], cactus thorn-like geometries [23], or variations in surface wettability [24, 25] have shown attractive droplet drainage features for laboratory-scale setups, but they lack scalability of mass production for industrial or community fog harvesting applications [26, 27]. Considering these factors, we argue that mass-producible interwoven metal meshes, like the one in Figure 1A, are scalable and structurally sound, and hence best suited for such applications.

It is important to note here that the horizontal fibers of such a woven mesh can offer resistance to droplet drainage [28], and therefore lead to serious clogging issues [2, 7]. Vertical elements within these meshes play a pivotal role in facilitating the drainage of accumulated fog droplets [29], thus posing no threat related to clogging. In this context, Shi et al. [30, 31] and Goswami et al. [32] have underscored the effectiveness of harp-like mesh – the mesh has only an array of vertical elements – particularly in moderate fog conditions. However, the harp-design is found to be less viable in scenarios involving heavy fogging or in tilted arrangements (as one requires in cooling tower fog harvesters [7]; see Figure 1B) due to their structural weaknesses and susceptibility to tangling [33]. The vulnerability of mesh to clog, as already mentioned, depends on the geometry of the mesh and its surface wettability (characterized by the apparent contact angle; see Figure 1A inset). This leaves rooms for mesh design improvement through appropriate choice of mesh wettability and geometry. To design a *'clog-proof'* interwoven metal mesh, it is logical to comprehend the behavior and morphology of deposited fog-water droplets during their evolution (see Figure 1C) on the mesh elements [34, 35]. A drop on the fiber of a fog harvesting mesh grows by consequent droplet coalescence and fog deposition, wherein it can assume either an axisymmetric barrel shape or an asymmetric clamshell shape (see Figure 1C) [35, 36, 37]. The droplet detaches from the fiber when its weight exceeds its adhesion force with the fiber [38, 39]. However, there is another possibility: the droplet may stretch transversely or laterally to reach neighboring mesh fibers and form capillary bridges, thereby leading to an increase of surface adhesion force and preventing detachment [40]. A droplet interacting with more than one fiber can therefore acquire larger volumes before it would detach [41], and hence accentuate the chance of clogging up the mesh pore. This would not only reduce the intended drainage of the collected fog-water (into the designated collector) but also increase aerodynamic resistance, thereby reducing total mesh efficiency.

While there have been multiple studies on the development of novel mesh designs or suitable surface modifications [42, 12, 2] to improve fog collection, few have investigated into the understanding of the interwoven mesh geometry parameters and droplet morphology interactions that lead to clogging [40]. Herein, we attempt to theoretically design the salient fog harvesting mesh parameters, viz., the fiber diameter and inter-fiber spacing, by drawing heavily from droplet morphology, such that a pendant droplet on a fiber does not grow large enough to contact the neighboring mesh fibers thereby ensuring a *clog-proof* design. We deconstruct the mesh into its individual fibers, thus allowing for a focused examination of a single pendant droplet morphology and its interaction with a horizontal element of the mesh. The intent here is to leverage the droplet morphology data to identify the conditions that lead to mesh pore clogging, and deduce from it the mesh design criteria, viz., the fiber diameter and pitch and its surface wettability, that would prevent such clogging. Thereafter, we calculate the critical shade coefficients – the maximum fraction of the projected mesh area that is obstructed by the solid fibers of the mesh – for such *clog-proof* meshes, and estimate the corresponding aerodynamic properties, and deposition efficiencies. Findings of the study serves as the baseline criteria for design of high-efficiency fog harvesting meshes.

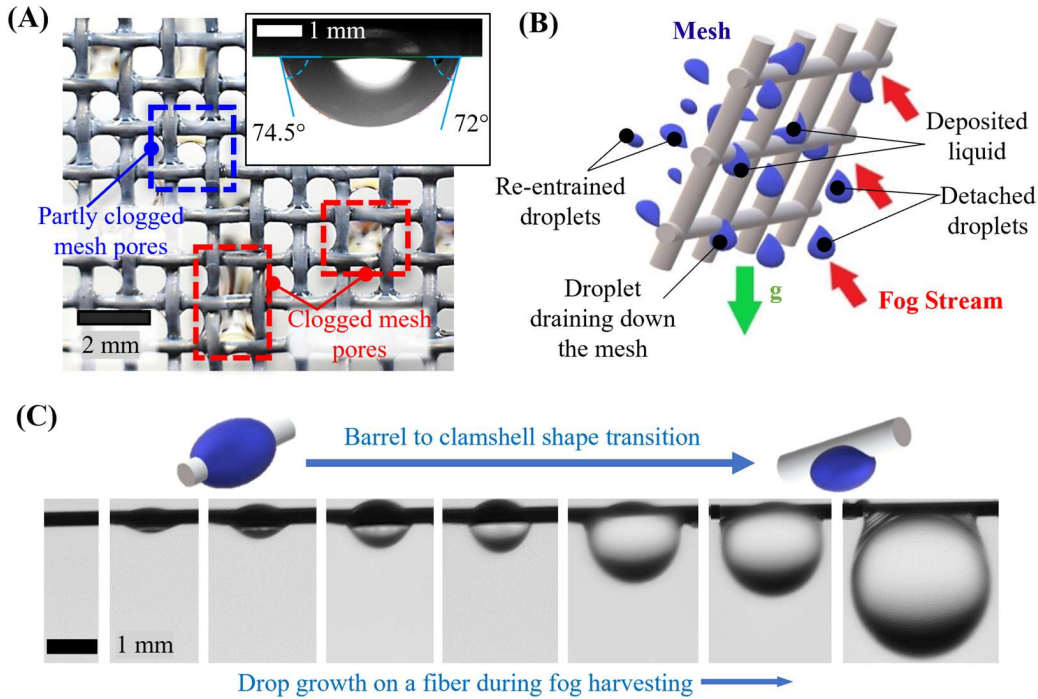
3

**Figure 1: (A)** Traditional interweaved metal meshes are prone to clogging of mesh pores. Mesh clogging has strong dependence on the apparent contact angle of the liquid droplet on the fiber surface (inset). **(B)** A typical mesh arrangement in cooling tower fog harvesters [3] showing deposited and detached droplets. **(C)** A growing droplet on a thin fiber can undergo dramatic morphological variations: from axisymmetric barrel shaped droplets to clamshell shapes.

## 2 Methodology

As already mentioned, the wettability of the mesh fibers, represented by its apparent contact angle ($\theta$), plays a strong role in determining the droplet morphology and whether it will clog the mesh under standard operation scenarios [43]. We investigate the droplet morphologies arising during fog harvesting, both experimentally and numerically, and identify the optimal design to avoid the formation of such clogging. Effects of varying fiber diameter and fiber surface wettability were also characterized to arrive at the design of *clog-proof* meshes.

### 2.1 *Experimental*

Experiments were carried out to characterize the morphology of water droplets on horizontally mounted mesh fibers through direct imaging. Fibers of three different wettability, viz., untreated fibers (control), superhydrophilic (SHPL), and hydrophobic (HPB) were chosen. For the control surfaces, stainless steel fibers (SS-304), with measured roughness ($R_a$) ~ 0.29 ± 0.11 μm and outer diameters varying from 0.25 – 6 mm were used to mimic the fibers of a fog harvesting mesh. The fiber diameters were measured with a digital screw gauge (Yuzuki, least count of 0.001 mm). All surface roughness measurements were carried out using a contact stylus profilometer (Landtek Instruments). All the fibers were cleansed via bath sonication (PCI Analytics), first in acetone and then in distilled water for 10 minutes. Thereafter, the apparent contact angle ($\theta$) measurements were performed using an optical goniometer (Holmarc), for sessile droplets in the clamshell shape, as depicted earlier in Figure 1A (inset) [44, 45]. Smooth aluminum cylinders of outer diameter 2 – 6 mm and $\theta$ ~ 71° ± 3° were turned SHPL by sandblasting, followed by etching in 3N hydrochloric acid (HCl) solution which creates micro-nano roughness on the surface. The SHPL surfaces were then passivated in boiling water for 30 minutes thereby creating stable hierarchical böhmite structures [46]. The fibers displayed a very low $\theta$ (< 5°) with a roughness of $R_a$ ~ 4.01 ± 0.22 μm [47]. The control fibers were turned HPB ($\theta$ ~ 104° ± 3°, measured roughness, $R_a$ ~ 0.58 ± 0.31 μm) by dip-coating with polydimethylsiloxane (PDMS): fibers were first ultrasonicated in acetone and water, dried and then dip-coated (NXT dip-KPM, Apex Instruments) in a PDMS solution (SYLGARD™ 184 Silicone Elastomer, Dow; premixed thoroughly with a crosslinking agent at a ratio of 10:1 by weight) at an axial draw-out velocity of 10 mm/s. The coated fibers were then cured in a hot-air convection oven at 120 °C for ~ 2 hours.

An experimental setup as shown in Figure 2A was developed for obtaining the droplet morphology and dimensions through direct imaging. A horizontally positioned three-jaw chuck was used to firmly hold onto the wettability-engineered cylindrical fibers. A flicker-free LED-powered (LT Max, GSVitec) white screen was used as a background for all imaging purposes. Live feed from a digital camera (Nikon D7200, with Nikkor AF-SDX 50 mm lens, mounted on 12 – 36 mm extension tubes) was stored onto a computer and viewed simultaneously on a monitor during experimentation. A micropipette (Thermo Scientific) was used to gently dispense liquid volumes on top of the fibers (see Figure 2B) in quanta of $V_{LD}$ = 1.0 μL (the least count of the dispenser).

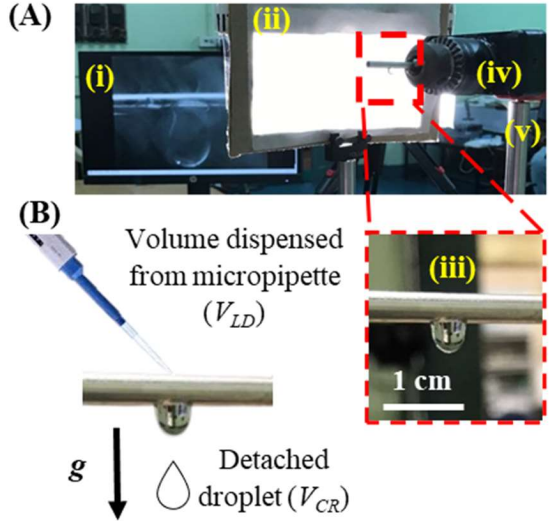

**Figure 2 (A)** Experimental setup with a computer display **(i)** and an illuminated white background **(ii)** to capture droplet morphology on a wettability-engineered metal fiber **(iii,** the zoomed-in section). The metal fiber was held by a horizontally placed three-way chuck **(iv)**, which in turn was affixed to a firm stand **(v)**. **(B)** Schematic of the arrangement for droplet volume addition using a micropipette. The volume was incremented, in steps of the least dispensable volume ($V_{LD}$) of the micropipette, until the droplet detached at $V_{CR}$.

After each droplet had coalesced with the previously deposited droplet, images were recorded, and the process of dispensing another droplet was repeated until the accumulated droplet detached from the fiber; the corresponding volume was termed the critical volume of droplet detachment ($V_{CR}$). Care was taken while dispensing the droplet, such that the pipette tip did not touch the fiber – the droplets were dispensed with minimum disturbance to the fiber – to avoid any external perturbations (e.g., wind shear, vibrations, etc.) and to minimize any unwanted dynamic behavior leading to droplet detachment. This technique is in-line with previously reported approaches adopted to replicate liquid deposition from fog impingement [48, 49].

## 2.2 Simulation

Surface Evolver (SE) simulations [50] were conducted with horizontal fibers of different diameters and wettability regimes, with droplets of different volumes deposited on them. The SE simulation implicitly assumes a smooth surface for the fibers, notwithstanding their micro- and nanoscale surface roughness features in reality. Keeping in view that surface roughness plays a key role in determining the apparent contact angle $\theta$ of water ($\gamma_{LV}$ = 72 mN/m) on a surface [51, 52, 46], experimentally recorded $\theta$ values from goniometer measurements were used as input in the SE simulations [53, 54]. The results from the SE simulations were compared with pre-existing data in the literature, and against Carroll's analytical expression for barrel-to-clamshell shape transitions [11, 55]. The versatility of the SE code is captured from the capability to model such transition phenomena. The governing equations and pertinent boundary conditions, the adaptive grid refinement and convergence criteria, and data processing for estimation of maximal droplet dimensions from the simulation results have been elucidated in the Electronic Supplementary Information (ESI) Section S1. The developed code was extended to accommodate for variations in fiber Bond number (*Bo*) arising from a change in fiber diameter or due to the gravitational influence. Thereafter, parametric variations were carried out to identify the conditions of the morphology transition for varying *Bo* scenarios. The minimum volume considered for observing the droplet morphology in SE was of the order of 0.1 μL. The effects of line tension prevalent for nanoscopic droplets was, therefore, neglected [56]. To further generalize our theory, the simulation results have also been compared with the experiments over a wide range of wettability for both superhydrophilic and hydrophobic fibers.

## 3  Results and Discussions

### *3.1  Barrel or clamshell shape?*

#### *3.1.1  Surface Evolver (SE) simulations*

As per the literature, a liquid droplet on a horizontal, cylindrical fiber can assume either an axisymmetric barrel shape or an asymmetric clamshell shape (see Figure 1C). For example, a barrel shape is energetically preferred in cases where liquid volume is high, and the sessile droplet contact angles (of the droplet on the fiber) are low. On the contrary, clamshell shapes are energetically stable at lower liquid volumes or for high sessile droplet contact angles [35]. However, in-between such regimes of absolute stability, there exist metastable regimes wherein both shapes emerge as energetically preferred. Factors like diameter and surface chemistry of the underlying fiber significantly influence the wetting properties, thereby playing a strong influence on the final droplet morphology on the fiber [35, 57]. For example, decreasing the reduced volume ($V_R = V/(r_f)^3$, where $r_f$ = the fiber radius and $V$ = the volume of the fiber-attached droplet) or rendering the surface hydrophobic ($\theta$ > 90°) results in a "roll-up" of the barrel to clamshell shape [34]. For low fiber Bond number ($Bo = \rho g r_f^2 / \gamma_{LV}$, where $r_f$ = the fiber radius, $\rho$ = liquid density, $g$ = acceleration due to gravity, and $\gamma_{LV}$ = surface tension at the air-water interface) scenarios, both these shapes can be energetically feasible [58].

To ascertain the regimes of droplet morphology and its dependence on $V_R$ and $\theta$, simulations were carried out in SE for drop-on-fiber systems for fixed $r_f$ (corresponding to fixed fiber *Bo*). For this purpose, an arbitrary volume of the drop was chosen, in any fixed geometry pertaining to either a stable barrel or clamshell shape. Thereafter, simulations were repeated with an increment or decrement of $\theta$ in steps of 1°) until the drop morphed into another stable geometry, or got detached from the fiber. A sequence of such simulations has been shown in Figure 3A, where the $\theta$ was incremented for fixed $V_R$ and fiber *Bo* and a transformation of the droplet from a barrel shape to a stable asymmetric clamshell shape was observed.

To demarcate the absolute stability regimes in barrel shape at a fixed $V_R$ and fiber *Bo*, SE simulations were carried out from an initially stable clamshell droplet at relatively higher liquid-fiber $\theta$ (> 30°). Thereafter, $\theta$ was continually decreased in steps of 1° and the simulation was repeated until the clamshell-shaped drops was no longer stable, and the drop morphed into a barrel shape. This transition marked the beginning of the absolute stability of the axisymmetric barrel shapes. All such transition points, wherein the drop morphed from a metastable clamshell into a stable barrel shape have been plotted in Figure 3B. Therefore, any drop-on-fiber system located to the left of the stability regimes shown in Figure 3B must be in the absolute barrel shape, whereas drops to the right represents a metastable clamshell/barrel shape regime (i.e., both barrel and clamshell shapes are energetically feasible). Furthermore, with an increase in fiber *Bo* (or increasing gravitational influence), the absolute stability regime in the barrel shape shifts leftward on the $V_R - \theta$ plane, indicating that greater surface tension forces would be required to maintain a stable barrel shape (Figure 3B). At greater fiber *Bo* (≥ 0.0044) this regime is only present at very low $\theta$ ($\leq 5°$). Interestingly, it may be pointed out here that under idyllic scenarios and in the absence of any perturbations/gravitational effects ($Bo = 0$) barrel shapes can grow indefinitely large. However, even the slightest gravitational effects or perturbations would morph it into an energetically preferred clamshell.

A similar methodology was implemented for establishing the absolute stability in clamshell regime for any particular fiber $Bo$. An axisymmetric barrel droplet of fixed $V_R$ was first simulated on a fiber, after which the $\theta$ was incremented in steps of 1° until the droplet transitioned into an asymmetric clamshell. Thereafter, such simulations were extended for varying $Bo$ to populate the morphology transition map diagram for stable clamshell geometry (Figure 3C). As explained earlier, any drop-on-fiber system to the left of such transition points would represent a metastable barrel/clamshell, whereas systems to the right of such transition curves represent absolute stability in the clamshell shape. Contrary to barrel-shaped droplets, clamshell droplets are found to be energetically more favored for increasing fiber $Bo$ (increasing gravitational influence). An example case to illustrate this simulation methodology implemented in estimating absolute stability regimes for a drop-on-fiber system at a fiber $Bo$ of 0.0002 is provided in the Electronic Supplementary Information (ESI) Section S1.2.

It is also interesting to note that, during an event of sustained droplet growth (e.g., via coalescence during fog harvesting or in condensation applications) or shrinkage (e.g., due to evaporation), the droplet shape can morph from one regime to the other (e.g., stable to meta-stable or vice versa), and then again back to the previous regime if the volume change continues further in the same direction (e.g., growth or shrinkage). Such phenomenon, termed as "re-entrant behavior" [59], has been highlighted both in Figures 3B and 3C with dashed lines. This behavior is observed during droplet growth on a surface having considerably low $\theta$ at higher fiber $Bo$. For example, if we consider the case of fiber $Bo = 0.0218$ in Figure 3C, the dashed line along $\theta \approx 15°$ indicates that a droplet exhibiting a stable clamshell shape for $V_R < 40$ shifts to the meta-stable regime for a higher value of $V_R$, where it may also remain in a barrel shape (provided, however, it originated from a barrel shape). On further growth of the droplet, corresponding to $V_R > 140$, the droplet regains the absolute stability regime wherein only clamshells are energetically favored. A similar re-entrant behavior (meta-stable to stable-barrel, and then again to a meta-stable regime upon progressively increasing $V_R$) is also observed for the case of fiber $Bo = 0.0044$ in Figure 3B, along the dashed line at $\theta \approx 4°$. This behavior is an outgrowth of the simultaneous interplay of the gravitational potential energy and the surface energy of the fiber and the liquid.

Reckoning the morphological phase diagram constituted by Figures 3B and 3C is particularly relevant for the fog harvesting scenarios. In reality, the droplet would start to grow on the fiber from a very low value of $V_R$, which would engender a stable clamshell morphology (see Figure 3C). At larger values of $V_R$ – the reduced volume increases progressively as the droplet grows because of subsequent fog deposition – the droplet morphology would be dictated both by the $Bo$ and $\theta$. Following observations of Figures 3B and 3C, the droplet would continue to remain in a clamshell morphology until the combination of $Bo$ and $\theta$ encroaches in the stable barrel morphology regime. It may be noticed from Figure 3C that even for the thinnest fiber ($r_f = 0.25$ mm) investigated here (see Section 2.1), under normal gravitational ($g = 9.81$ m/s$^2$, corresponding $Bo$ of 0.00212, which would lie on the right-hand side of the $Bo = 0.0044$ line) scenarios, a pendant clamshell morphology is always energetically preferred over an axisymmetric barrel (see Section S3.0 in ESI). Hence, in the following sections describing the experimental observations, and in further investigations for the design of *clog-proof* meshes, we limit our discussions only to clamshell shaped drops.

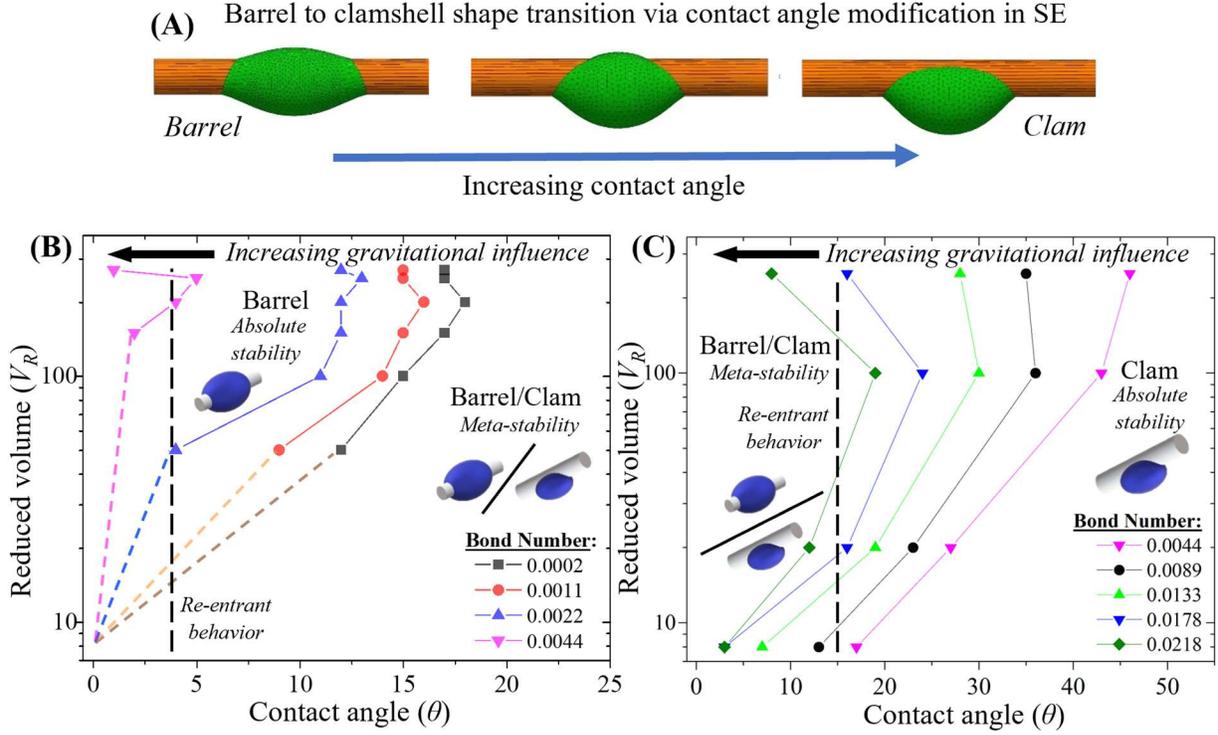

**Figure 3: (A)** Estimation of stable droplet morphology regimes from the SE simulations. Regime maps of droplet morphology on $V_R - \theta$ plane for different $Bo$: **(B)** lines demarcating the absolute barrel stability (left of each curve for the pertinent $Bo$) and barrel/clamshell metastable regimes; **(C)** lines demarcating the absolute clamshell stability (right of each curve for the pertinent $Bo$). Example cases of re-entrant behavior of the droplet from one stability regime to another and back have been marked with the dashed vertical lines in the regime plots.

*3.1.2 Experimental investigations into droplet morphology*

As explained earlier, experiments were undertaken on wettability engineered fibers to evaluate droplet morphologies and to attest to the simulation results obtained from SE. A series of images to capture the droplet growth on varying fiber diameters have been provided in Figure 4. It was observed that there is significant difference in the droplet morphology for variations in fiber diameter or wettability. As is evident from Figure 4A, droplets on SHPL fibers had a greater lateral spread compared to their vertical extent, while droplets on HPB fibers had smaller contact areas. Such a morphology can be explained theoretically by taking the surface energies into consideration: SHPL fibers have greater surface energy, therefore sessile droplet geometries result in maximized contact area and vice versa. Furthermore, the extent of the droplet morphology was also limited by its critical volume of detachment ($V_{CR}$), wherein the droplet detached from the fiber. As the volume of the droplet is increased, the lateral extent of droplets on SHPL fibers kept on increasing till they reach a maximum, as opposed to their HPB counterparts (wherein, extension of the vertical hang is favored compared to the increase in contact area of the drop-on-fiber system, Section S1.3 in ESI). We characterize the change of such extents by describing morphological parameters and using them to characterize *clog-proof* meshes in the following sections.

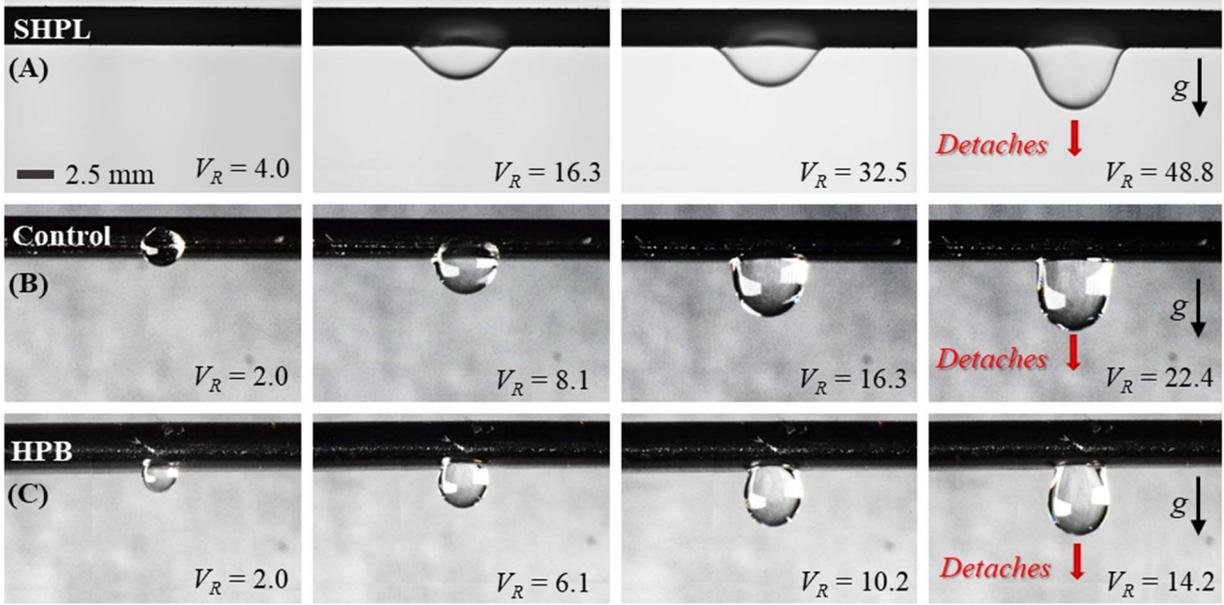

**Figure 4:** Clamshell droplet morphologies (for varying $V_R$) on wettability engineered fibers, as observed during experimentation. Droplets on SHPL fibers **(A)**, were seen to spread laterally and had a greater lateral extent. Droplets on control fibers (untreated SS) showed no preference in change of lateral or vertical spread on increase of droplet volume **(B)**, whereas droplets on HPB surfaces minimized their contact area while extending vertically with increase of volume **(C)**. Comparisons between lateral and vertical extent of varying drop-on-fiber morphologies have been carried out in the next section. All experiments were carried out until the droplet reached the critical volume of detachment ($V_{CR}$, as seen in the last frame from each sequence), any volume addition after this would lead to droplet detachment. All images have the same scale bar, while each image show a fiber of diameter 2.7 mm.

Furthermore, for all the experimental trials conducted throughout this study, the control fiber ($\theta \approx 70°$) diameters were varied from 0.25 mm to 6 mm, which corresponds to fiber $Bo$ of 0.00212 and 1.22625 [35]. Again, for experiments in the super-hydrophilic regime ($\theta < 5°$), surface-treated aluminum fibers had diameter variations from 2 mm to 6 mm, corresponding to $Bo$ variations from 0.13625 to 1.22625. From earlier stability regime plots in Figure 3, it is evident that for both these sets of experiments, barrel-shaped droplets would not be energetically preferred for any $V_R$. Similarly, barrel-shaped droplets for fiber radii in the order of ~ 1 mm (corresponding $Bo$ of 0.0340), in hydrophobic regimes ($\theta > 90°$) are impossible under normal gravitational ($g = 9.81$ m/s$^2$) scenarios. Therefore, and as explained earlier, realistic barrel-shaped droplets do not appear both during numerical simulations and in experimental investigations of collected droplets on fog harvesting meshes (Section S3.0 in ESI), and understanding morphologies of clamshell droplets hold key in design of *clog-proof* meshes.

### 3.2 Design and characterization of "clog-proof" meshes

#### 3.2.1 Droplet maximal dimension

As evident from the SE simulations and the experiments described in the previous sections (refer to Figure 3A or Figure 4), a clamshell droplet growing on a cylindrical fiber extends its vertical hang ($H$) and its lateral spread ($W$, measured at the base or the girth of the droplet, whichever was greater), until the droplet detaches from the fiber because of its weight. These characteristic dimensions have been described in Figure 5A, wherein the consistency between the experimentally observed and simulated droplet morphology have also been demonstrated for stainless steel control (SS 304, $\theta \sim 70°$) fibers of diameter 1.27 mm (top image, droplet volume 10 μL, $V_R$ = 39.1) and 4 mm (bottom image, droplet volume 45 μL, $V_R$ = 5.6). It is seen from Figure 5A that the droplet hanging from the 1.27 mm fiber has $W > H$, while that from the 4 mm fiber exhibits $W < H$. It was also observed (See ESI Figure S3 B) that while $H$ kept on increasing until detachment, the lateral spread only increased up to a certain maximum ($W_{max}$) and then decreases as the "detachment neck" forms in the droplet. Therefore, while $H_{max}$ occurs just prior to the detachment (i.e., $V_R = V_{CR}$), $W_{max}$ is observed at $V_R < V_{CR}$. Such a behavior ensues from the variation of the self-adjusting surface adhesion forces with the changing droplet footprint as it attempts to balance the gravitational influence on the droplet.

To examine the relative extents of $H$ and $W$ of a clamshell-shaped pendant droplet from the fiber, our SE simulations and experiments were further extended to describe the $W$ and $H$ values for different droplet-fiber $\theta$ and fiber diameters ($2r_f$ = 0.4, 1.27, 2.0, 4.0, 5.0 and 6.0 mm, where $r_f$ is the fiber radius). For SE simulations, droplet morphologies for varying $\theta$ = 15°, 30°, 70°, 110°, and 160° were estimated, while for experiments the SHPL ($\theta < 5°$), control ($\theta \sim 70°$), and HPB ($\theta = 104°$) surfaces were chosen. For both SE simulations and experiments, stable droplet morphologies and the corresponding $H$ and $W$ values were noted for different values of $V_R$ – starting from zero to $V_{CR}$ with incremental steps of $V_{LD}$. The largest values of $W$ and $H$ assumed by the droplet in the volume range ($0 \leq V_R \leq V_{CR}$) are recorded from each data set as $W_{max}$ and $H_{max}$, respectively. The methodology of extracting the $H_{max}$ and $W_{max}$ from the SE simulation and the experimentally obtained images is elaborated in the ESI Section 1.3. These values of $W_{max}$ and $H_{max}$, obtained over the entire parametric regime of $\theta$ and $r_f$ are used to arrive at the fog harvesting mesh design criteria as described below.

As shown in Figure 5B, a droplet on the fiber of a mesh could always interact with its neighboring fibers to form capillary bridges, which would clog the mesh pore and impact fog capture adversely. Hence, for a given fiber radius of the mesh that has square-shaped pores, the pitch (distance between two fiber centerlines) needs to be greater than the maximal dimensions of the droplet spread or hang (viz., $H_{max}$ and $W_{max}$ respectively) to ensure that it never clogs. The underlying assumption is that the fog-water droplet grows symmetrically from a central point marked by the green arrows in Figure 5B. Although, in reality, fog droplets grow at multiple locations on the horizontal fiber of a mesh cell, this assumption serves the purpose without eliciting much error because of a practically observed phenomenon. It was noticed from the experiments that fog droplets growing asymmetrically, or near the corners of a mesh pore touches the nearest vertical fiber and drains down, leaving little chance of clogging the mesh pore. On the contrary, a droplet that grows from the middle of the mesh pore is at a greater risk of clogging the mesh pore as it will almost simultaneously touch vertical fibers on either side (Figure 5C and ESI Section S3.0). Therefore, we treat this configuration of fiber-attached droplet as the representative case to design clog-proof meshes.

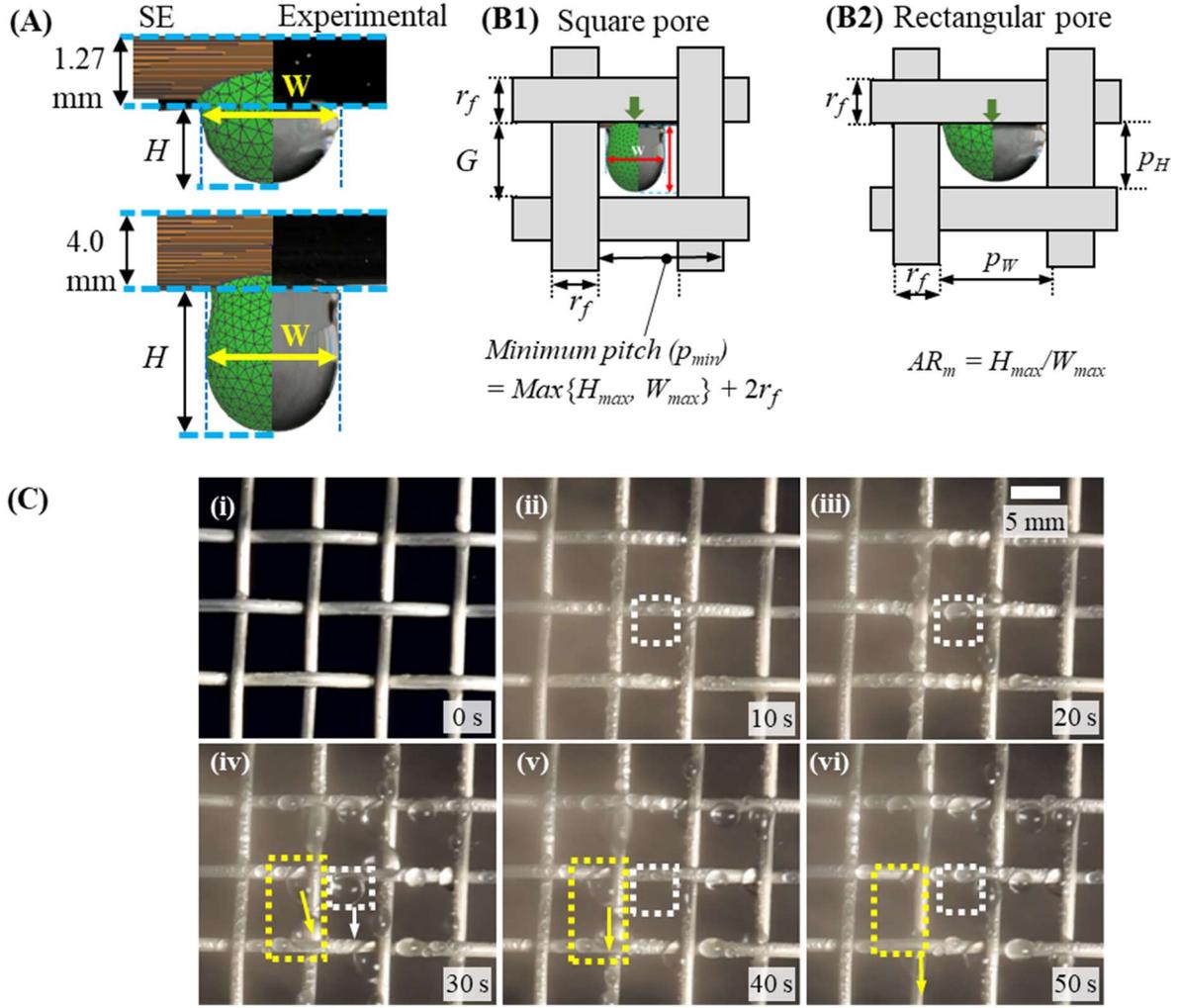

**Figure 5: (A)** Measurement of maximal width ($W_{max}$) and maximal vertical hang ($H_{max}$) in droplets hanging from 1.27 mm (top, droplet volume 10 μL) and 4 mm (bottom, droplet volume 45 μL) fibers (stainless steel 304, $\theta \sim 70°$). Each figure comprises of two halves: the left half (colored in green) corresponds to SE simulations, while the other half is an experimental image. **(B)** Rationale for the design of *clog-proof* mesh with square pores **(B1)** and rectangular pores **(B2)**: The mesh pitch should be such that a symmetrically growing droplet should not touch any other fiber, either on the side or at the bottom, as it grows until its detachment. The mesh pore dimensions, viz., the pore height and width have been marked as $p_H$ and $p_W$ respectively. **(C)** Fog droplets growing from the middle of the mesh pore have a greater probability of clogging the mesh (marked in dashed white box), while droplets near a vertical fiber (marked in dashed yellow box) can drain down after contact with the nearest vertical fiber).

To further generalize the design of *clog-proof* fog harvesting meshes, both square and rectangular mesh pores have been considered. It is evident from Figure 5 B1 that the pitch ($p$) between the fibers for a square mesh is the sum of the mesh pore side ($G$, corresponding either of $p_W$ or $p_H$ for a square-pored mesh) and the fiber diameter, i.e.,

$$p = G + 2r_f \qquad (1)$$

Once maximal droplet dimensions viz., $H_{max}$ and $W_{max}$, had been evaluated for a specific combination of fiber radius and wettability, the minimum mesh pore dimensions $p_{min}(\theta, r_f)$ for a square-pore fog harvesting mesh were set equal to maximum of either $H_{max}$ or $W_{max}$ added to the mesh-fiber diameter, i.e.,

$$p_{min}(\theta, r_f) = Max\{H_{max}, W_{max}\} + 2r_f \qquad (2)$$

This design approach for square-pored meshes would ensure that the droplet detaches before contacting any neighboring fibers. The maximal droplet dimension, evaluated for the first term on the right-hand side of Equation 2, is shown in Figure 6A as a function of the fiber diameter for different levels of fiber wettability as considered for SE simulations or measured (for experiments) $\theta$.

For rectangular pore mesh designs (as previously illustrated in Figure 5 B2) the height- and width-wise dimensions, denoted as $p_H$ and $p_W$, respectively, are not necessarily equal, and for designing such mesh, one should also keep in mind the aspect ratio ($AR$), i.e., the ratio of the height- and width-wise droplet dimensions. The $AR$ estimated for the maximal dimensions of the droplet (for a particular fiber diameter and level of surface wettability), have been defined as the maximal aspect ratio and denoted as ($AR_m = H_{max}/W_{max}$). For a mesh with rectangular pores, the minimum value of the longer pitch (measured either height- or width-wise) should be linked with the maximal droplet dimension as

$$p_{min,longer}(\theta, r_f) = Max\{H_{max}, W_{max}\} + 2r_f, \qquad (3a)$$

while the minimum value of the shorter pitch may be evaluated as

$$p_{min,shorter}(\theta, r_f) = \frac{Max\{H_{max}, W_{max}\}}{F} + 2r_f \qquad (3b)$$

where $F = AR_m \quad \forall \; \{H_{max} > W_{max}\};$ \qquad (3c)

$\qquad = 1/AR_m \quad \forall \; \{H_{max} < W_{max}\};$ and

$\qquad = 1 \qquad \forall \; \{H_{max} = W_{max}\}$

Therefore, for rectangular mesh design, both the droplet maximal dimension (i.e., $Max\{H_{max}, W_{max}\}$) and the droplet aspect ratio $AR_m$ are needed to ascertain the minimum pitches in the two orthogonal directions. While the former is already available in Figure 6A, the $AR_m$ values have been plotted in Figure 6B against different fiber radii and for different wettability levels. Furthermore, a series of images in Figure 6C (taken during the experiments) show this variation of $AR$, at $V_{CR}$, for droplets on unmodified stainless steel control fibers.

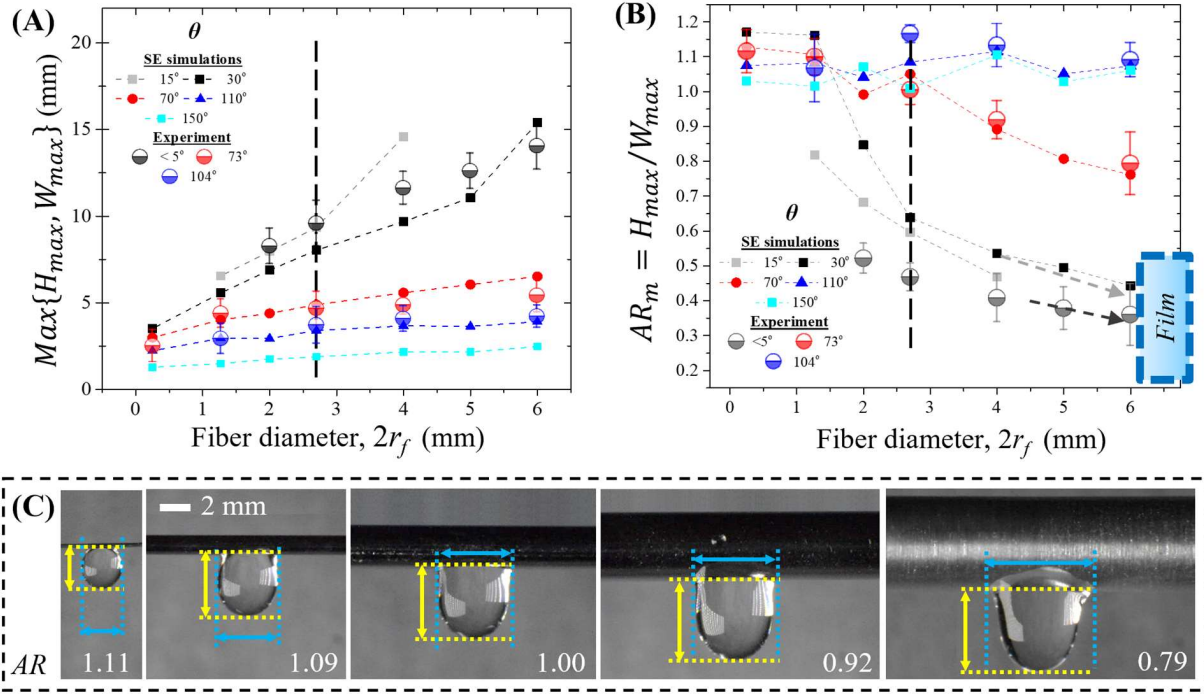

**Figure 6**: Variation of **(A)** the maximal droplet dimension ($Max\{H_{max}, W_{max}\}$) with the fiber radius, for the estimation of the minimum mesh pitch (as per Equation 2 or 3), and **(B)** the corresponding pore dimension maximal aspect ratio ($AR_m = H_{max}/W_{max}$, for the estimation of the minimum mesh pitch of a rectangular mesh as per Equation 3), plotted as functions of the fiber diameter for different mesh wettability. Legend: Filled symbols with dotted lines denote SE simulation results, half-filled circles denote the experimental data. Vertical dotted lines in both figures, mark the capillary length scale of water (~ 2.7 mm). **(C)** A series of experimental images showing the $AR=H/W$ calculated at the respective $V_{CR}$ for different control fiber diameters ($\theta = 73°$); $H$ and $W$ are marked in yellow and blue, respectively.

The maximal droplet dimension (Figure 6A) and droplet aspect ratio (Figure 6B) data show that both the SE simulations (filled symbols with dotted lines) and the experimental runs (half-filled circles) exhibit similar trend of variation with increasing $r_f$. The vertical dotted line in both the figures mark the capillary length scale of water (~2.7 mm). While the maximal droplet dimension (and hence the minimum mesh pitch length) is found to increase with the fiber diameter (Figure 6A), trends in variation of $AR_m$ are seen to be different on either side of the capillary length scale. For fiber diameters smaller than the capillary length (i.e., on the left side of the dashed line in Figure 6B), $AR_m$ is seen to converge near unity for a wide range of variation in wettability. This indicates that square shaped mesh pores are best suited for small fiber diameters. The final geometry of such *clog-proof* meshes would however depend on the wettability of the mesh fiber, and the *clog-proof* nature would rely on the minimum mesh pitch length as elucidated in Equations 2 and 3. Furthermore, for hydrophobic meshes, square mesh pores with appropriate minimum mesh pitch length appears to be *clog-proof*, since $AR_m \sim 1$ for almost all fiber diameters with $\theta = 110°$ or $150°$ (from the SE simulations) and $\theta = 104°$ (from the experimental measurements).

Therefore, it is evident from Figures 6A and 6B that the mesh-pore dimensions $p_H$ and $p_W$ are physical parameters entirely dependent on the drop-on-fiber morphology, the fiber diameter, and its wettability. While it is important in design of rectangular-pored *clog-proof* meshes, $AR_m$ can also be interpreted as the ratio of the gravitational and surface-adhesion length scales acting on the droplet. A low $AR_m$ (< 1.0) indicates greater lateral spread of the droplet throughout its growth (i.e., stronger influence of surface adhesion between the droplet and the fiber). On the other hand, a high $AR_m$ (> 1.0) value will imply that the droplet-fiber contact width is less, and gravitational influence is more predominant. This interpretation is further invigorated by the fact that hydrophilic fibers ($\theta$ < 90°) are observed to have decreasing $AR_m$ values with increasing fiber radii. The increasing lateral spread characterized by the decreasing $AR_m$ is prominent from the experiments and simulations involving very low $\theta$ (< 15°). For such highly wettable surfaces, an increase in fiber diameter sharply increases the maximal lateral spread (thus decreasing $AR_m$) and the droplet eventually spreads as a film, as indicated at the bottom right corner of the plots in Figure 6B. For SE simulations, runs with large fiber diameters (5 mm and 6 mm), did not to converge for very low $\theta$ (< 15°), as the droplet kept on spreading, forming a liquid film on the underside of the fiber. This observation also concurred with the experiments, where the contact line could no more be observed for superhydrophilic fibers, but the droplet would spread to form a film. We hypothesize that such film formation happens as capillary forces between the fiber and the droplet, at higher fiber diameter and lower $\theta$, are sufficiently larger than gravitational forces and maximum energy minimization is obtained by assuming a film shaped morphology in place of a droplet. Our present study therefore excludes this film regime, since the mesh pitch dimensions, we intend to investigate here are in the millimetric range, which is at least an order of magnitude higher than the film thickness.

### *3.3 'Clog-proof' mesh selection criteria and efficiencies*

Once the droplet morphology parameters, necessary to define the geometry of *clog-proof* meshes, are determined and relevant mesh design parameters viz., $r_f$, $p_H$ and $p_W$ calculated, it becomes pertinent to evaluate the fog-collection performances of such optimally designed meshes. Furthermore, the estimation of fog harvesting mesh efficiencies require knowledge of shade coefficient, which depends on the fiber radius and the mesh pore geometry. The shade coefficient encompasses all parameters relevant to the mesh pore geometry, including the mesh pore aspect ratio ($AR_m$), and therefore estimation of mesh efficiencies can be carried out for all *clog-proof* meshes described herein. However, while the design criteria for both square-pore and rectangular-pore fog harvesting *clog-proof* mesh geometries have been specified earlier, only investigation of fog-capture performance of square-pored meshes have been taken up. The efficiencies of rectangular pored meshes can be evaluated in similar ways as described in the following sections.

As fog harvesting meshes intercept fog-laden wind, to collect droplets via subsequent deposition and coalescence on its fibers, it also offers a significant pressure drop to the flow. The deposited droplet eventually drains down the mesh for collection and later use. The total capture efficiency of such meshes are determined from the product of aerodynamic, deposition, and drainage efficiencies [3,7]. We use the aerodynamic theory for the fog laden airflow through the mesh and the deposition dynamics of the fog droplets in the mesh that is designed for *clog-proof* operation following the minimum-pitch criteria described in the previous section. However, estimation of the drainage efficiency for such meshes is beyond the scope of the present study and may be taken up as a future exercise.

*3.3.1   Aerodynamics of flow through a mesh*

Fog capture meshes intercepting fog laden winds, present an interesting, but complicated flow over several bluff bodies, which causes a pressure drop across the mesh. The fluid dynamics of such an interception via a porous blunt mesh can be considered analogous to flow across a porous media, and modeled by considering three regions of interest during the flow as depicted in Figure 7A – *Region 1* wherein the upstream fog flow gets deflected owing to the obstruction imposed by the mesh in its path, *Region 2* denotes the location just after the net, while *Region 3* is located at the far downstream. As shown in Figure 7A, a stream tube (denoted by the blue-dotted lines) of cross section $A_1$, smaller than the area $A$ of the fog mesh, is defined to quantify the volume flow rate of the fog-laden flow through the mesh. Following the principle of continuity, for a fog stream approaching the mesh with free-stream velocity $u$, it can be shown that the cross-section areas $A_1$ and $A$ are related with upstream velocity, $u$, and velocity of the fog-stream just after interaction with the mesh, $w$ (as depicted in Figure 7A) as

$$A_1 u = Aw \text{ or}, \frac{A_1}{A} = \frac{w}{u} = u^* \tag{4}$$

Here, $u^*$ is defined as the "velocity ratio" for the given fog-flow conditions. For better estimation of the fog mesh aerodynamic efficiency, we resort to approximations as previously implemented by Koo [60] and Steiros [61]:

- Fog flow in region 1 is outside the stream tube encompassing the mesh and are inviscid and irrotational (potential flow approximations)
- In region 2, the fog flow follows the streamlines and is inviscid.
- Fog flow in region 3 is viscous, and characterized by significant mixing with the outer flow.
- There is significant mixing of the wake with outer flow in the interface between regions 2 and 3 which is independent of the induced velocity $w$.

The "fog interception ratio", $\varphi = (A_1/A)$ (also referred in earlier studies as the "filtered fraction") represents the fraction of the unperturbed flow reaching the mesh without getting diverted. From this, the aerodynamic efficiency ($\eta_a$, the fraction of the unperturbed oncoming flow which is directed toward the projected solid area of the mesh) can be estimated as by $\eta_a = \varphi \cdot SC$ where $SC$ is the shade coefficient of the mesh [20]. Therefore, the aerodynamic efficiency can be expressed both as a function of the velocity ratio and the fog-interception ratio, by combining the definition of $\eta_a$ with Equation 5, such that,

$$\eta_a = \frac{w}{u} SC = u^* \cdot SC = \varphi \cdot SC \tag{5}$$

Therefore, for estimation of the fog-interception ratio ($\varphi$), pressure drops at the microscale (for inviscid flows through a porous substrate) are compared with the resulting pressure drop at the macroscale arising due to interaction of the fog-stream with a blunt body (herein the fog-mesh) at higher Reynolds number [62]. This comparison leads to a direct definition of the fog-interception ratio ($\varphi$) as shown in Equation 6,

$$\varphi = \frac{w}{u} = u^* = \frac{A_1}{A} = \sqrt{\frac{C_D}{k}} \tag{6}$$

where $C_D$ and $k$ are the drag and pressure drop coefficients arising from the fluid mechanical analysis of the problem at the macroscale and the microscale approach to the problem respectively [62]. Furthermore, such coefficients in inviscid flows through a solid blunt porous body for potential flow approximations have been extensively explored in earlier research by Koo [60] and Steiros [61]. $k$ and $C_D$ are directly related to the velocity ratio ($u^*$) and the shade coefficient ($SC$) as expressed in equations 7 and 8, respectively.

$$k = \left(\frac{1}{(1-SC)^2} - 1\right) - \frac{4}{3}\frac{(1-u^*)^3}{u^{*2}(2-u^*)^2} \tag{7}$$

$$C_D = \frac{4}{3}\frac{(1-u^*)(2+u^*)}{(2-u^*)} \tag{8}$$

Using equations 5 – 8, the velocity ratio can be implicitly solved for $SC \in [0,1]$, and may be expressed as a function of the shade coefficient or, $u^* = f(SC)$, using a linear regression fit (Equation 9a). This explicit form may, in turn, be used to estimate the aerodynamic efficiency as a function of $SC$ in a fog harvesting mesh (Equation 9b).

$$u^* = 1.09 - 1.05 SC, \ R^2 = 0.99 \tag{9a}$$

$$\eta_a = 1.09 SC - 1.05 SC^2 \tag{9b}$$

Figure 7B shows the aerodynamic efficiency of square-pore fog harvesting meshes of varying fiber diameter and wettability, wherein the $SC$ was chosen using the criteria of minimum mesh pitch $p_{min}(\theta, r_f)$ based on the maximal droplet dimensions (Figure 6A), to realize the *clog-proof* nature of the mesh. Therefore, for a given fiber diameter and wettability,

$$SC = \left[1 - \frac{2r_f}{p_{min}(\theta, r_f)}\right]^2 \tag{10}$$

Each curve for *clog-proof* mesh $SC$ in Figure 7B – drawn at a specified wettability values – divides the graph into two regions. The region upwards of each curve (i.e., at a greater $SC$, implying that the mesh pitch for a given fiber diameter is smaller than that obtained from Equation 3) indicates a regions of mesh design prone to clogging. Whereas the region below the optimal $SC$ curve implies that the mesh pitch is larger than the limit set by Equation 3, and hence the mesh will remain clog-free. To describe the overall performance of the mesh, both in terms of its aerodynamic efficiency and ability to avoid clogging, Figure 7B is also shaded with the color map contours of aerodynamic efficiencies, as computed from Equation 9b. For better clarity, the aerodynamic efficiency - $SC$ curve is also shown by the side in Figure 7B(ii), wherein the aerodynamic efficiency is seen to peak ($\eta_{a,max}$ = 28.31 %) at an optimum of $SC_{opt}$ = 0.52.

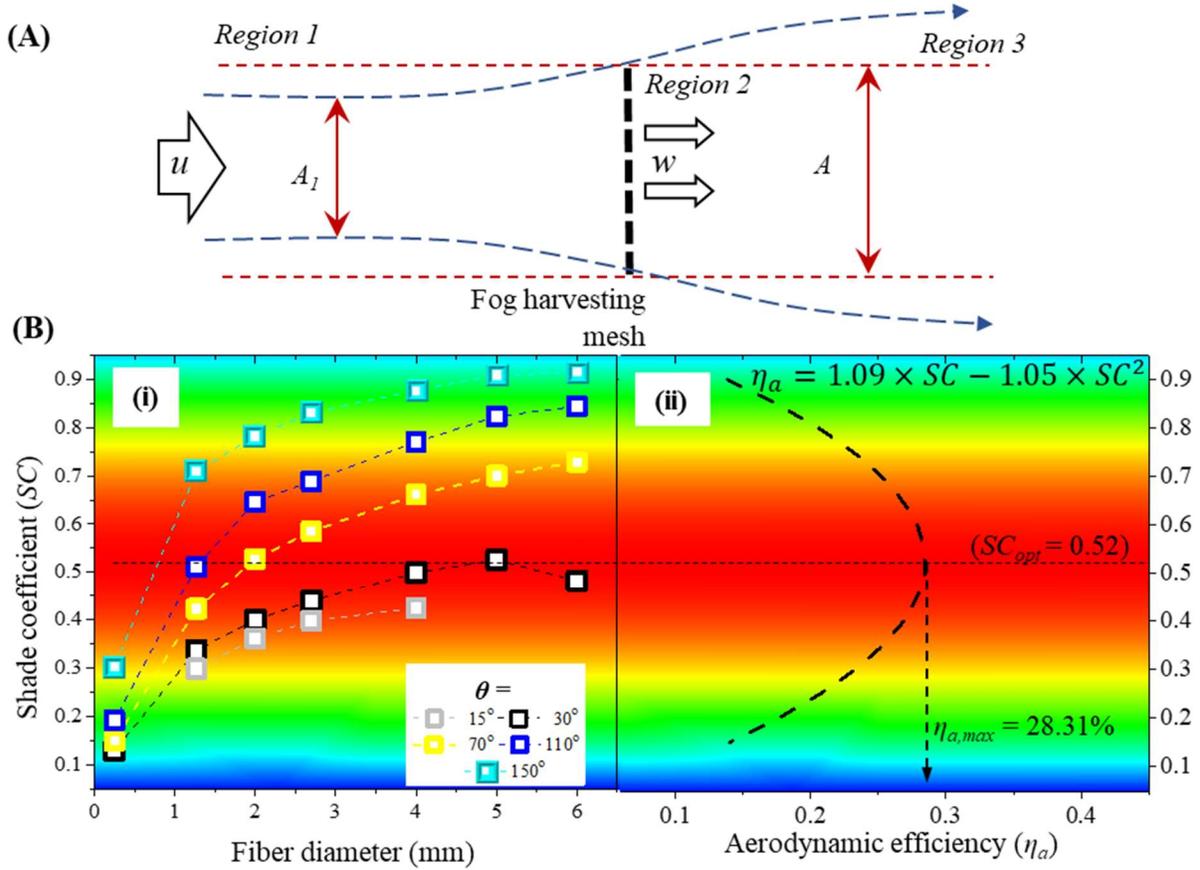

**Figure 7: (A)** Schematic of flow through a mesh placed in the path of fog laden wind (after Steiros [61]). Flow in *Region 1* deviates and goes around the mesh; fog within the upstream cross-sectional area $A_1$ (less than the projected mesh area) interacts with the fog mesh. *Region 2* occurs just after the mesh, while *Region 3* is located at far downwind where mixing of flows from *Regions 1* and *2* occur. **(B-i)** Variation of the *clog-proof SC* values with the mesh fiber diameter for different fiber wettability. The *clog-proof SC* is calculated corresponding to the minimum mesh pitch lengths for square-pore fog meshes as suggested in Figure 6A. Regions above each curve (corresponding to a greater *SC* for a particular fiber diameter and wettability) are prone to clogging. Background color map denotes the corresponding aerodynamic efficiency as evaluated from **(B-ii)**, which describes the variation of aerodynamic efficiency (also the black dashed curve) with the *SC* for square-pore fog-harvesting meshes. The color map background in **(B-ii)** is the same as that in **(B-i)**, for easy comparison of $\eta_a$ with *clog-proof SCs*.

Figure 7 indicates that, besides keeping the mesh design *clog-proof*, it is further possible to optimize the mesh geometry to maximize the aerodynamic efficiency close to the aforementioned value of $\eta_{a,max}$. This is illustrated in Figure 8A by showing the *clog-proof* design regimes of mesh fibers of three representative wettabilities (viz., $\theta$ = 30°, 70° and 150°) on the *SC* vs fiber diameter plane. The shaded regions (cross-hatched in respective colors) below each curve designate the design regime of *clog-proof* operation, while unshaded parts above the curves denote regimes of potential clogging. The rationale of mesh design would, therefore, be to remain within the shaded region (of the respective mesh-wettability), and have *SC* as close as possible to $SC_{opt} = 0.52$. It may be seen from Figure 8A that a highly hydrophobic ($\theta =150°$) mesh, which has a relatively smaller $p_{min}(\theta, r_f)$ may be designed as *clog-free* and also with $SC = SC_{opt}$ (= 0.52) for any fiber diameter greater than 0.8 mm. Therefore, the mesh is expected to remain clog-free and yield $\eta_{a,max}$ = 28.31 % for $2r_f > 0.8$ mm. For thinner fibers (i.e., $2r_f < 0.8$ mm), the maximum allowable *SC* for clog-proof operation will have to be lower, and hence the efficiency will be lower than $\eta_{a,max}$. For fibers with $\theta =70°$, only meshes having $2r_f > 2.0$ mm may be designed with the optimum *SC*. Similarly, for a wettable mesh ($\theta=30°$), $p_{min}(\theta, r_f)$ is much larger, and only meshes having $2r_f \sim$ 5 mm can be designed with $SC = SC_{opt}$. Therefore, the corresponding plots of maximum achievable $\eta_a$ values of the *clog-proof* meshes of different fiber diameters and for varying levels of fiber-wettability have been shown in Figure 8B. It is imperative from Figure 8B that $\eta_a$ increases with the fiber diameter and attains $\eta_{a,max}$ beyond a certain value. It also transpires from Figure 8B that with smaller fiber diameter meshes, it is better to have a hydrophobic mesh to maintain both the attributes of high aerodynamic efficiency and *clog-proof* nature.

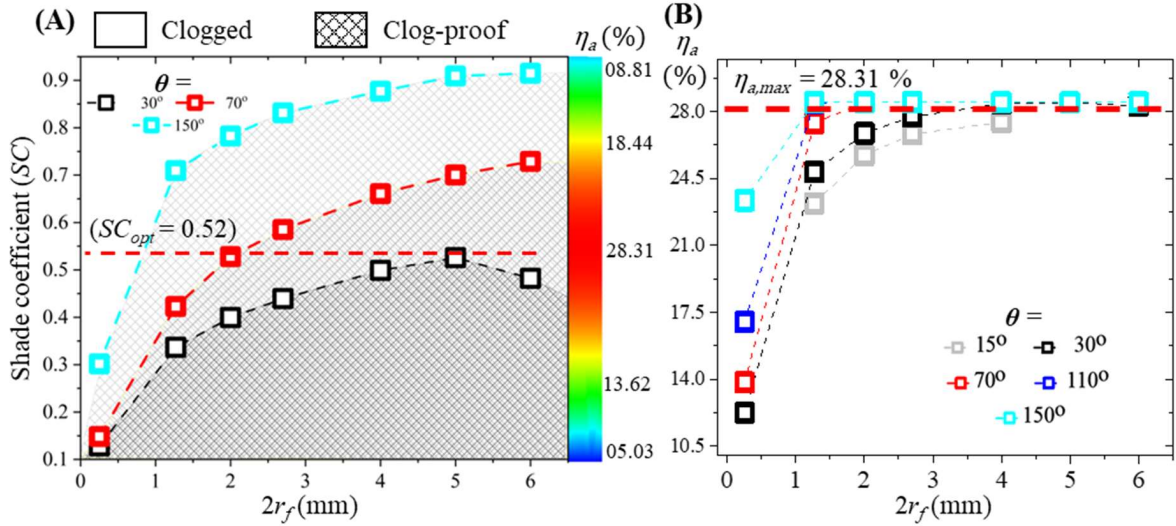

**Figure 8**: **(A)** Optimal shade coefficients for varying fiber diameter and wettability. The optimal mesh *SC* is calculated corresponding to *clog-proof* mesh pitch lengths from Figure 6A. Regions above each curve (for a specific level of fiber wettability) are prone to clogging (trends with only 3 representative fiber-wettability are shown for clarity). **(B)** The resulting maximum achievable aerodynamic efficiency for *clog-proof* square meshes as a function of fiber diameter for different fiber wettabilities.

Therefore, from the results we conclude that *clog-proof* meshes with lower fiber diameters results in lower aerodynamic efficiency and provides greater impedance to the flow of fog and vice versa. Furthermore, variation of the fog-mesh fiber wettability was also seen to have a considerable impact on the design and resulting aerodynamics of *clog-proof* meshes. While it might be *wrongly* inferred here that clog proof meshes with larger fiber diameters are best suited for fog harvesting, this is not always true as the effect of deposition efficiency must be taken into consideration before commenting on the overall mesh efficiency.

### 3.3.2 Deposition and total efficiencies

Having established a high aerodynamic efficiency is not the only precondition of high overall capture efficiency. The fog droplets should also have a high deposition rate (and deposition efficiency) on the fiber meshes. Langmuir and Blodgett [63] in their seminal work developed a correlation for deposition of small droplets headed towards an infinitely long cylinder, thereby demonstrating the effects of the inertial and viscous forces. The capture efficiency of such a fog particle from a flow field is determined by its Stokes number (*St*), defined as the ratio of the characteristic time of the particle ($t_p$) to the characteristic time of flow ($t_f$) around the collector (herein, a smooth cylindrical fog-mesh fiber element). The stopping time of the particle is conventionally computed using Stokes linear drag law and in such reversible Stokesian flow regimes wherein the inertial forces are small and the particles can be assumed to be perfect solid spheres, the drag force on a fog particle can be estimated as $F_D = 6\pi R_{fog} \mu_{air} u$, where $R_{fog}$ is the radius of the fog particle, $\rho_{fog}$ the density of fog water, $\mu_{air}$ the dynamic viscosity and $u$ the flow velocity of the carrier medium (herein air), respectively. For a particle moving initially with the free stream velocity, the deceleration can therefore be simply estimated from Newton's second law as

$$a = \frac{F_D}{m} = \frac{6\pi R_{fog} \mu_{air} u}{(4/3)\pi R_{fog}^3 \rho_{fog}} = \frac{9\mu_{air} u}{2 R_{fog}^2 \rho_{fog}} \tag{11}$$

The characteristic stopping time ($t_p$) after which the fog particle gets deposited on the fiber, may be estimated by momentum balance of a fog particle, such that,

$$t_p = \frac{u}{a} = \frac{2 R_{fog}^2 \rho_{fog}}{9\mu_{air}} \tag{12}$$

The characteristic time of flow ($t_f$) can be defined as the ratio of the length scale of the fog flow around a mesh fiber to the velocity of the free stream ($t_f = 2r_f/u$). Plugging the values for the characteristic times, the Stokes number (*St*) can be represented in terms of the mesh fiber Reynolds number ($Re_{fiber} = \rho_{air} u_{slip} (2r_f)/\mu_{air}$) as

$$St = \frac{2}{9} Re_{fiber} \frac{\rho_{fog}}{\rho_{air}} \left(\frac{R_{fog}}{2r_f}\right)^2 \tag{13}$$

The fog droplet deposition efficiency ($\eta_d$) for millimeter sized mesh-pores can be calculated directly by using the empirical expression that is widely acceptable in the literature [2, 63, 64] as

$$\eta_d = \frac{St}{(\pi/2)+S} \tag{14}$$

However, such estimation of deposition efficiency is limited only to very low fog particle Reynolds number, defined as $Re_{fog} = 2\rho_{air} u R_{fog}/\mu_{air}$. The average fog droplet diameter ($2R_{fog}$) and has been assumed to be ~ 5 µm throughout the remainder of this study [2]. The estimation of deposition efficiencies at higher $Re_{fog}$, calls for the definition of a generalized Stokes number ($St_g$). Similar proposition for a generalized *St* had been earlier put forth by Israel and Rosner [65], to account for any non-Stokesian drag on a particle in a stream. According to the revised analysis (see ESI S2.0 for the derivation), the deposition efficiency $\eta_d$ may be expressed in terms of the generalized Stokes number $St_g$ as

$$\eta_d^{-1} \cong 1 + 1.25 \left(St_g - \frac{1}{8}\right)^{-1} - 1.4 \times 10^{-2} \left(St_g - \frac{1}{8}\right)^{-2} - 0.508 \times 10^{-4} \left(St_g - \frac{1}{8}\right)^{-3} \tag{15}$$

for $St_g > 0.14$. Thus, for all our fog deposition efficiency estimations, we resort to using the Israel and Rosner generalized deposition efficiency (Equation 15) for any $St_g > 0.14$, while using the Langmuir-Blodgett [63] equations with $St_g$ (Equation 14) for $St_g \leq 0.14$. Figure 9A plots the different deposition efficiencies for variations in fiber radii for different realistic sets of fog free-stream velocities ranging up to 10 m/s. A higher fog stream velocity is seen to result in a higher inertial deposition of the fog droplets, thereby leading to an increase in the deposition efficiency for all fiber diameters (Figure 9A). However, smaller fiber diameter result in better deposition than that observed on larger ones. This may be attributed to the fact that a slender fiber gives rise to a sharper deviation of the streamlines as the flow encounters it, leading to a larger inertial effect as the particles attempt to flow past the fibers. For further characterization of overall efficiencies of the *clog-proof* meshes, a free fog stream flow velocity of 5 m/s is chosen, since it is typical in industrial cooling tower fog harvesting applications.

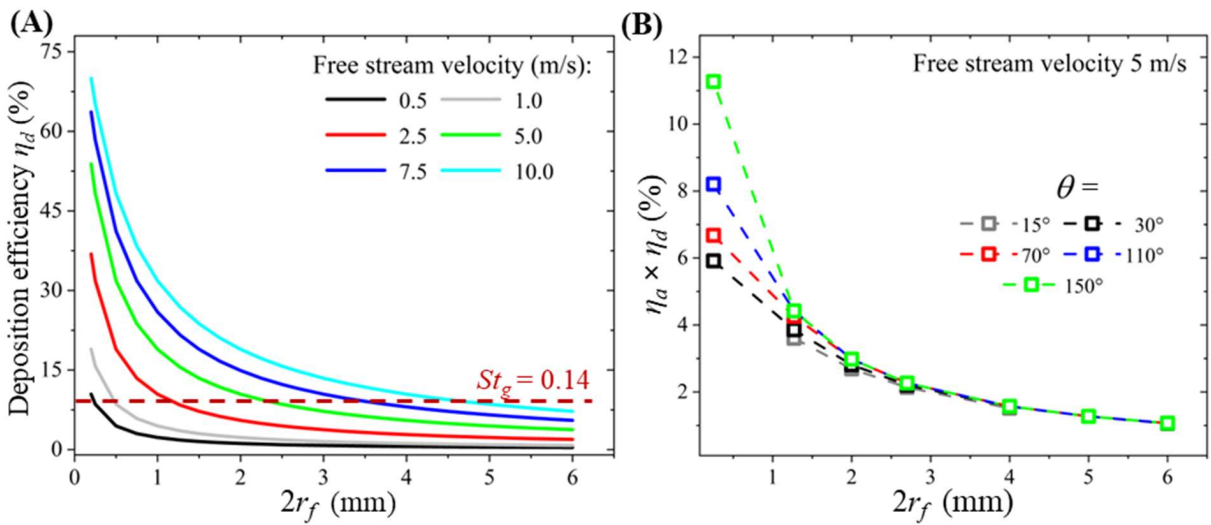

**Figure 9: (A)** Variation of deposition efficiencies ($\eta_d$) as function of fiber diameters at different fog flow velocities as evaluated from Equation 14 for $St_g \leq 0.14$ and from Equation 15 for $St_g > 0.14$. $\eta_d$ resulting from $St_g = 0.14$ has been marked with a dashed red line. **(B)** Variation of ($\eta_d \times \eta_a$) at different optimal *clog-proof* shade coefficients for different mesh fiber diameters and wettability conditions. Total efficiencies have been estimated for a representative fog flow velocity of 5 m/s.

Reckoning the optimum shade coefficients pertaining to the maximum aerodynamic efficiency for design of *clog-proof* meshes corresponding to their wettability and fiber diameters (Figure 6B) and the deposition efficiencies from Figure 9A are combined to deduce the overall efficiency $\eta_d \times \eta_a$ of the *clog-proof* fog harvesters, which is reported in Figure 9B. The overall product of aerodynamic and deposition efficiencies is observed to decrease with increase in fiber diameters for designed *clog-proof* meshes. This decrease in the overall efficiency is in stark contrast to variation of aerodynamic efficiency with fiber diameter variation. As can be understood from the variation of both the efficiencies, deposition efficiency plays a stronger role and dominates to determine the final trend of the overall efficiency. Although, there is an increase in aerodynamic efficiency with increasing fiber radii (as discussed in the last section), this is not sufficient to compensate for the decrease in deposition efficiency. It is also noted that for fiber radii below ~2.7 mm (the capillary length-scale of water), the fog collection efficiency improves with increased hydrophobicity of the fiber, while retaining the *clog-proof* of the mesh. However, for larger mesh fiber diameter, the mesh fiber wettability is found to play very little role. This arises from the fact that the maximum achievable $\eta_a$ of the *clog-proof* design of the mesh became almost insensitive to the mesh wettability for fiber diameter exceeding ~2.7 mm.

The article provides the basis for the design of a fog harvesting mesh that estimates the optimal mesh pitch for a combination of mesh fiber diameter and the fiber wettability to ensure clog-free operation and maximizes the overall collection efficiency. While thin hydrophobic fibers are found perform efficiently due to high deposition efficiencies in optimal *SC clog-proof* designs, it is important to keep in mind that they tend to suffer from re-entrainment and premature drainage in fog harvesting applications [7]. Incorporation of drainage efficiencies and re-entrainment factors to better realize such *clog-proof* meshes, remains a future direction. Although the current study does not take into consideration the events of premature dripping and flow-induced re-entrainment, it attempts to ameliorate the occurrence of mesh clogging, which is a major component attributing to the fog harvester performance degradation during field deployment.

## 4   Conclusions and Outlook

We have investigated the development of a *clog-free* mesh design by incorporating droplet morphology maximal parameters ($H_{max}$ or $W_{max}$) into the geometry of fog harvesting meshes. Although numerous forms of deposited droplet configurations can exist on the mesh, we have simplified our problem by deconstructing the mesh and looking at the droplet growth and interaction on a single smooth cylindrical horizontal mesh fiber. First, the quasi-static growth and morphology evolution of a water droplet on a cylindrical fiber of given diameter and surface wettability was investigated experimentally and through Surface Evolver simulations. Thereafter, the maximal height and spread of the droplets, as they hang from the horizontal fiber, until they grow large enough to be shed by gravity, were noted, and used to estimate the minimum mesh pitch that would ensure clog-free operation in fog harvesting meshes. The estimated minimum mesh pitch was then used to specify minimum shade coefficients (*SC*) of square mesh pore fog harvesters. This optimal *SC* was further used to calculate aerodynamic, deposition, and overall mesh efficiencies in typical industrial scale cooling tower fog harvesting applications. We find that for fog harvesting meshes designed at the *clog-proof* condition, hydrophobic fibers ($\theta > 90°$) with diameters below the capillary length scale, outperform other fibers in terms of overall efficiency. For *clog-proof* meshes with fiber diameters above the capillary length scale, mesh wettability was seen to play very little role in determining the overall efficiency. Further, the overall efficiency was seen to decrease with increase of fiber diameter. Such a trend can be attributed to the fact that the optimally-designed *clog-proof* meshes, with a square pore, have a constant aerodynamic efficiency corresponding to an optimal *SC* of 0.52 at larger diameters and the trend in overall efficiency is directly influenced by the decreasing deposition efficiency of larger fibers.

It is also important to note the limitations of the present study. While the role of the underlying surface texture (and pertinent physicochemical properties) were captured in our numerical simulations by using goniometric measurements of apparent contact angles [46, 57], the model does not account for contact line-pinning effects or frictional forces at the contact line [66]. Keeping in view that in realistic situations, mass-manufactured, wettability-engineered fog-harvesting meshes would entail surface imperfections at the micro-scale and even surface impurities (e.g., oil, volatile organic compounds, chemicals, scales and dust), investigation into the role of contact-line pinning in clogging of fog harvesting meshes may be taken up as a future exercise, using the current results as the baseline data. Similar exploits into wettability engineered mesh durability, reliability of performance and maintenance aspects also remain to be investigated under realistic scenarios [67, 68] to estimate the impact of ageing on the clog-free operation of the mesh. Also, it is worth noting that the actual metal mesh is often formed by weaving, where the gaps between the crossed mesh wires can create additional capillary effect, leading to pinning of droplets and partial clogging as shown in Figure 1A. The present model may be extended in future to investigate the capillary forces arising from the intersection of fibers [28] under different degrees of fiber wettability, intersection angle and inter-fiber spacing. The present work also upholds the importance of droplet morphology dimensions during its evolution – from impaction and coalescence with neighboring drops to consequent detachment – in design of *clog-proof* fog harvesters and their performance, while opening a new strategy for design of better fog harvesters.

## 5 Supporting Information

- Droplet growth and drainage in fog harvesting scenarios in non-optimized interwoven meshes (video).
- Droplet growth and evolution in morphology as fog droplets coalesce and grow on a SS-304 fiber ($2r_f$ = 0.25 mm, θ ~ 73°) till detachment (video).
  - No barrel-shaped droplets were observed, as had been predicted earlier from the SE simulations in the main text.
  - Optimizing mesh pore geometry to minimize clogging based on maximal dimensions, as calculated during such evolution of the droplet.
- Surface Evolver simulations; Validation of SE simulations; Methodology for barrel/clam regime estimation and convergence criterion; Data processing for estimation of maximal droplet dimensions; Derivation of generalized Stokes number ($St_g$); (PDF)

## 6 Acknowledgement

Authors gratefully acknowledge the funding from DST-SERB, India through Project No. CRG/2019/005887. AM thankfully acknowledge the assistantship from JU-RUSA. Help from Rupam Mahanta in editing of the *ESI* videos is also gratefully acknowledged.

## 7 Declaration of competing interests

AM, AD, AD and RG are the inventors on a patent application titled "Never-clog' mesh designs for capture and separation of a dispersed phase from a flowing fluid stream and systems thereof" related to this work filed with Indian patent office (application no: 202231045629, filed on 10[th] August 2022).

## 8 References

[1] Connor, R. The United Nations world water development report 2015: water for a sustainable world (Vol. 1). UNESCO publishing. **2015**.
[2] Park, K. C.; Chhatre, S. S.; Srinivasan, S.; Cohen, R. E.; McKinley, G. H. Optimal design of permeable fiber network structures for fog harvesting. Langmuir. **2013**, 29(43), 13269-13277.
[3] Ghosh, R.; Ray, T. K.; Ganguly, R. Cooling tower fog harvesting in power plants–A pilot study. Energy. **2015**, 89, 1018-1028.

[4] Schemenauer, R. S.; Bignell, B.; Makepeace, T. Fog collection projects in Nepal: 1997 to 2016. Proceedings of the 7th International Conference on Fog, Fog Collection and Dew, Wroclaw, Poland. **2016**, 187-190.

[5] Schemenauer, R. S.; Cereceda, P.; Osses, P. FogQuest: Fog water collection manual. **2005**.

[6] Ghosh, R.; Ganguly, R. Harvesting water from natural and industrial fogs—opportunities and challenges. Droplet and spray transport: paradigms and applications. **2018**, 237-266.

[7] Ghosh, R.; Ganguly, R. Fog harvesting from cooling towers using metal mesh: Effects of aerodynamic, deposition, and drainage efficiencies. Proc. Inst. Mech. Eng., Part A. **2020**, 234(7), 994-1014.

[8] Rothman, T.; Ledbetter, J. O. Droplet size of cooling tower fog. Environ. Lett. **1975**, 10(3), 191-203.

[9] Stechkina, I. B.; Kirsch, A. A.; Fuchs, N. A. Studies on fibrous aerosol filters—iv calculation of aerosol deposition in model filters in the range of maximum penetration. Ann. Occup. Hyg. **1969**, 12(1), 1-8.

[10] Lorenceau, É.; Clanet, C.; Quéré, D. Capturing drops with a thin fiber. J. Colloid Interface Sci. **2004**, 279(1), 192-197.

[11] Datta, A.; Mukhopadhyay, A.; Dutta, P. S.; Saha, A.; Datta, A.; Ganguly, R. Droplet detachment from a horizontal fiber of a fog harvesting mesh. Fluid Mechanics and Fluid Power (Vol. 2). FMFP 2021. Lecture Notes in Mechanical Engineering, Springer, Singapore. **2021**, 485-490.

[12] Ghosh, R.; Patra, C.; Singh, P.; Ganguly, R.; Sahu, R. P.; Zhitomirsky, I.; Puri, I. K. Influence of metal mesh wettability on fog harvesting in industrial cooling towers. Appl. Therm. Eng. **2020**, 181, 115963.

[13] Ang, B. T. W.; Zhang, J.; Lin, G. J.; Wang, H.; Lee, W. S. V.; Xue, J. Enhancing water harvesting through the cascading effect. ACS Appl. Mater. Interfaces. **2019**, 11(30), 27464-27469.

[14] Park, J. K.; Kim, S. Three-dimensionally structured flexible fog harvesting surfaces inspired by Namib Desert Beetles. Micromachines. **2019**, 10(3), 201.

[15] Raut, H. K.; Ranganath, A. S.; Baji, A.; Wood, K. L. Bio-inspired hierarchical topography for texture driven fog harvesting. Appl. Surf. Sci. **2019**, 465, 362-368.

[16] Zhang, L.; Wu, J.; Hedhili, M. N.; Yang, X.; Wang, P. Inkjet printing for direct micropatterning of a superhydrophobic surface: toward biomimetic fog harvesting surfaces. J. Mater. Chem. A. **2015**, 3(6), 2844-2852.

[17] Zhai, L.; Berg, M. C.; Cebeci, F. C.; Kim, Y.; Milwid, J. M.; Rubner, M. F.; Cohen, R. E. Patterned superhydrophobic surfaces: toward a synthetic mimic of the Namib Desert beetle. Nano Lett. **2006**, 6(6), 1213-1217.

[18] Ju, J.; Bai, H.; Zheng, Y.; Zhao, T.; Fang, R.; Jiang, L. A multi-structural and multi-functional integrated fog collection system in cactus. Nat. Commun. **2012**, 3(1), 1247.

[19] Lee, J.; So, J.; Bae, W. G.; Won, Y. The design of hydrophilic nanochannel-macrostripe fog collector: Enabling wicking-assisted vertical liquid delivery for the enhancement in fog collection efficiency. Adv. Mater. Interfaces. **2020**, 7(11), 1902150.

[20] de Dios Rivera, J. Aerodynamic collection efficiency of fog water collectors. Atmos. Res. **2011**, 102(3), 335-342.

[21] Chakrabarti, U.; Paoli, R.; Chatterjee, S.; Megaridis, C. M. Importance of body stance in fog droplet collection by the Namib desert beetle. Biomimetics. **2019**, 4(3), 59.

[22] Yu, Z.; Yun, F. F.; Wang, Y.; Yao, L.; Dou, S.; Liu, K.; Jiang, L.; Wang, X. Desert beetle-inspired superwettable patterned surfaces for water harvesting. Small. **2007**, 13(36), 1701403.

[23] Ju, J.; Xiao, K.; Yao, X.; Bai, H.; Jiang, L. Bioinspired conical copper wire with gradient wettability for continuous and efficient fog collection. Adv. Mater. **2013**, 25(41), 5937-5942.

[24] Breuer, C.; Cordt, C.; Hiller, B.; Geissler, A.; Biesalski, M. Using Paper as a Biomimetic Fog Harvesting Material. Adv. Mater. Interfaces. **2024**, 2301048.

[25] Kennedy, B. S.; Boreyko, J. B. Bio-Inspired Fog Harvesting Meshes: A Review. Adv. Funct. Mater. **2023**, 2306162.

[26] Fessehaye, M.; Abdul-Wahab, S. A.; Savage, M. J.; Kohler, T.; Gherezghiher, T.; Hurni, H. Fog-water collection for community use. Renewable Sustainable Energy Rev. **2014**, 29, 52-62.

[27] Bhushan, B. Design of water harvesting towers and projections for water collection from fog and condensation. Philos. Trans. R. Soc., A. **2020**, 378(2167), 20190440.

[28] Weyer, F.; Duchesne, A.; Vandewalle, N. Switching behavior of droplets crossing nodes on a fiber network. Sci. Rep. **2017**, 7(1), 13309.

[29] Gilet, T.; Terwagne, D.; Vandewalle, N. Droplets sliding on fibres. Eur. Phys. J. E: Soft Matter Biol. Phys. **2010**, 31, 253-262.

[30] Shi, W.; Anderson, M. J.; Tulkoff, J. B.; Kennedy, B. S.; Boreyko, J. B. Fog harvesting with harps. ACS Appl. Mater. Interfaces. **2018**, 10(14), 11979-11986.

[31] Shi, W.; van der Sloot, T. W.; Hart, B. J.; Kennedy, B. S.; Boreyko, J. B. Harps enable water harvesting under light fog conditions. Adv. Sustainable Sys. **2020**, 4(6), 2000040.

[32] Goswami, S.; Sidhpuria, R. M.; Khandekar, S. Effect of Droplet-Laden Fibers on Aerodynamics of Fog Collection on Vertical Fiber Arrays. Langmuir. **2023**, 39(50), 18238-18251.

[33] Shi, W.; De Koninck, L. H.; Hart, B. J.; Kowalski, N. G.; Fugaro, A. P.; van der Sloot, T. W.; Ott, R. S.; Kennedy, B. S.; Boreyko, J. B. Harps under heavy fog conditions: superior to meshes but prone to tangling. ACS Appl. Mater. Interfaces. **2020**, 12(42), 48124-48132.

[34] Adam, N. K. Detergent action and its relation to wetting and emulsification. J. Soc. Dyers Colour. **1937**, 53(4), 121-129.

[35] Eral, H. B.; de Ruiter, J.; de Ruiter, R.; Oh, J. M.; Semprebon, C.; Brinkmann, M.; Mugele, F. Drops on functional fibers: from barrels to clamshells and back. Soft Matter. **2011**, 7(11), 5138-5143.

[36] Carroll, B. J. The accurate measurement of contact angle, phase contact areas, drop volume, and Laplace excess pressure in drop-on-fiber systems. J. Colloid Interface Sci. **1976**, 57(3), 488-495.

[37] McHale, G.; Newton, M. I.; Carroll, B. J. The shape and stability of small liquid drops on fibers. Oil Gas Sci. Technol. **2001**, 56(1), 47-54.

[38] Saha, A.; Datta, A.; Mukhopadhyay, A.; Datta, A.; Ganguly, R. Time-Dependent Droplet Detachment Behaviour from Wettability-Engineered Fibers during Fog Harvesting. Fluid Mechanics and Fluid Power (Vol. 5). FMFP 2022. Lecture Notes in Mechanical Engineering, Springer, Singapore. **2024**, 463-471.

[39] Chou, T. H.; Hong, S. J.; Liang, Y. E.; Tsao, H. K.; Sheng, Y. J. Equilibrium phase diagram of drop-on-fiber: coexistent states and gravity effect. Langmuir. **2011**, 27(7), 3685-3692.

[40] Park, J.; Lee, C.; Lee, S.; Cho, H.; Moon, M. W.; Kim, S. J. Clogged water bridges for fog harvesting. Soft Matter. **2021**, 17(1), 136-144.

[41] Amrei, M. M.; Venkateshan, D. G.; D'Souza, N.; Atulasimha, J.; Tafreshi, H. V. Novel approach to measuring the droplet detachment force from fibers. Langmuir. **2016**, 32(50), 13333-13339.

[42] Sontag, D. S.; Saylor, J. R. An experimental study of the collection of fog droplets using a mesh fabric: Possible application to cooling towers. J. Energy Resour. Technol. **2016**, 138(2), 024501.

[43] Seo, D.; Lee, J.; Lee, C.; Nam, Y. The effects of surface wettability on the fog and dew moisture harvesting performance on tubular surfaces. Sci. Rep. **2016**, 6(1), 1-11.

[44] Mukhopadhyay, A.; Pal, A.; Sarkar, S.; Megaridis, C. M. Laser-Tuned Surface Wettability Modification and Incorporation of Aluminum Nitride (AlN) Ceramics in Thermal Management Devices. Adv. Funct. Mater. **2024**, 2313141.

[45] Datta, A.; Singh, V. K.; Das, C.; Halder, A.; Ghoshal, D.; Ganguly, R. Fabrication and characterization of transparent, self-cleaning glass covers for solar photovoltaic cells. Mater. Lett. **2020**, 277, 128350.

[46] Sarkar, S.; Roy, T.; Roy, A.; Moitra, S.; Ganguly, R.; Megaridis, C. M. Revisiting the supplementary relationship of dynamic contact angles measured by sessile-droplet and captive-bubble methods: Role of surface roughness. J. Colloid Interface Sci. **2021**, 581, 690-697.

[47] Das, C.; Gupta, R.; Halder, S.; Datta, A.; Ganguly, R. Filmwise condensation from humid air on a vertical superhydrophilic surface: Explicit roles of the humidity ratio difference and the degree of subcooling. J. Heat Transfer. **2021**, 143(6), 061601.

[48] Zhou, F.; Zhuang, D.; Lu, T.; Ding, G. Observation and modeling of droplet shape on metal fiber with gravity effect. Int. J. Heat Mass Transfer. **2020**, 161, 120294.

[49] Hu, H.; Lai, Z.; Hu, C. Droplet shedding characteristics on metal fibers with different wettability and inclined angles. Int. J. Refrig. **2023**, 130, 271-277.

[50] Brakke, K. A. The surface evolver. Experimental mathematics. **1992**, 1(2), 141-165.

[51] Wenzel, R. N. Surface roughness and contact angle. J. Phys. Chem. **1949**, 53(9), 1466-1467.

[52] Cassie, A. B. D.; Baxter, S. Wettability of porous surfaces. Trans. Faraday Soc. **1944**, 40, 546-551.

[53] Gennes, P. G.; Brochard-Wyart, F.; Quéré, D. Capillarity and wetting phenomena: drops, bubbles, pearls, waves. Springer New York. **2004.**

[54] Sinha Mahapatra, P.; Ganguly, R.; Ghosh, A.; Chatterjee, S.; Lowrey, S.; Sommers, A. D.; Megaridis, C. M. Patterning wettability for open-surface fluidic manipulation: fundamentals and applications. Chem. Rev. **2022**, 122(22), 16752-16801.

[55] Carroll, B. J. Equilibrium conformations of liquid drops on thin cylinders under forces of capillarity. A theory for the roll-up process. Langmuir. **1986**, 2(2), 248-250.

[56] Law, B. M.; McBride, S. P.; Wang, J. Y.; Wi, H. S.; Paneru, G.; Betelu, S.; Ushijima, B.; Takata, Y.; Flanders, B.; Bresme, F.; Matsubara, H.; Takiue, T. Aratono, M. Line tension and its influence on droplets and particles at surfaces. Prog. Surf. Sci. **2017**, 92(1), 1-39.

[57] Sarkar, S.; Gukeh, M. J.; Roy, T.; Gaikwad, H.; Bellussi, F. M.; Moitra, S.; Megaridis, C. M. A new methodology for measuring solid/liquid interfacial energy. J. Colloid Interface Sci. **2023**, 633, 800-807.

[58] Berthier, J.; Brakke, K. A.; Berthier, E. Open microfluidics. John Wiley & Sons. **2016.**

[59] Jiang, Y.; Machado, C.; Park, K. C. K. From capture to transport: A review of engineered surfaces for fog collection. Droplet. **2023**, 2(2), e55.

[60] Koo, J.-K.; James, D. F. Fluid flow around and through a screen. J. Fluid Mech. **1973**, 60(03), 513.

[61] Steiros, K.; Hultmark, M. Drag on flat plates of arbitrary porosity. J. Fluid Mech. **2018**, 853, R3.

[62] Azeem, M.; Guérin, A.; Dumais, T.; Caminos, L.; Goldstein, R. E.; Pesci, A. I.; de Dios Rivera, J.; Torres, M.J.; Wiener, J.; Campos, J.L.; Dumais, J. Optimal design of multilayer fog collectors. ACS Appl. Mater. Interfaces. **2020**, 12(6), 7736-7743.

[63] Langmuir, I.; Blodgett, K. A mathematical investigation of water droplet trajectories (No. 5418). Army Air Forces Headquarters, Air Technical Service Command. **1946.**

[64] Shahrokhian, A.; Feng, J.; King, H. Surface morphology enhances deposition efficiency in biomimetic, wind-driven fog collection. J. R. Soc., Interface. **2020**, 17(166), 20200038.

[65] Israel, R.; Rosner, D. E. Use of a generalized Stokes number to determine the aerodynamic capture efficiency of non-Stokesian particles from a compressible gas flow. Aerosol Sci. Technol. **1982**, 2(1), 45-51.

[66] McHale, G.; Gao, N.; Wells, G. G.; Barrio-Zhang, H.; Ledesma-Aguilar, R. Friction coefficients for droplets on solids: the liquid–solid Amontons' laws. Langmuir. **2022**, 38(14), 4425-4433.

[67] Showket, J.; Majumder, S.; Kumar, N.; Sett, S.; Mahapatra, P. S. Fog harvesting on micro-structured metal meshes: Effect of surface ageing. Micro Nano Eng. **2024**, 22, 100236.

[68] Ghosh, R.; Baut, A.; Belleri, G.; Kappl, M.; Butt, H. J.; Schutzius, T. M. Photocatalytically reactive surfaces for simultaneous water harvesting and treatment. Nat. Sustain. **2023**, 6(12), 1663-1672.